\documentstyle[twoside,fleqn,espcrc2]{article}

\newcommand{\AmS}{{\protect\the\textfont2
  A\kern-.1667em\lower.5ex\hbox{M}\kern-.125emS}}
\newcommand{\ignore}[1]{}
\newcommand{\smfrac}[2]{{\textstyle{{#1}\over{#2}}}}

\title{Extrapolation  Methods for the
       Dirac Inverter in Hybrid Monte Carlo\thanks{
       Talk presented by A.R. Levi}}

\author{R. C. Brower, A.R. Levi\address{Department of Physics,
        Boston University, 590 Commonwealth Ave.,
        Boston, MA 02215, USA}
        and
        K. Orginos\address{Department of Physics,
        Brown University, Providence, RI 02912, USA}}

\begin{document}

\begin{abstract}

In Hybrid Monte Carlo(HMC) simulations for full QCD,
the gauge fields evolve smoothly as a function of
Molecular Dynamics (MD) time.  Thus we investigate
improved methods of estimating the trial solutions
to the Dirac propagator as superpositions of the
solutions in the recent past.  So far our best
extrapolation method reduces the number of Conjugate
Gradient iterations per unit MD time by about
a factor of 4.  Further improvements
should be forthcoming as we further exploit the information of
past trajectories.
\end{abstract}

\maketitle

\section{INTRODUCTION}

The inclusion of internal fermion loops in the vacuum of QCD
is a major
challenge. The present state of the art for generating full QCD
configurations
is the so called Hybrid Monte Carlo algorithm  which uses
Molecular Dynamic evolution in a ``fifth time'' coordinate t.
The Hamiltonian
for this evolution is
\begin{equation}
         S = \smfrac{1}{2} Tr~P^2 + S_g(U) +
           \varphi^\dagger [ M^\dagger M ]^{-1}\varphi,
\end{equation}
where $P_\mu(x)$ are the angular momenta conjugate to
the gauge fields
$U_\mu(x)$, $S_g(U)$ is the pure gauge action,
$M(U)$ is the Dirac matrix and
$\varphi$ is the pseudofermion field. In our discussion the
precise form of
the gauge action is not important. What is relevant is the
need to accurately
integrate the equations of motion, calculating the
force on $U$ due to the
pseudofermions at each time step.  This requires solving,
over and over again,
the linear equation,
\begin{equation}
      A(t)~ \chi(t) ~=~ \varphi,
\label{eq:linear}
\end{equation}
where $A(t)\equiv M(U)^\dagger M(U)$ and $\chi(t)$
is the solution of the
inverted Dirac operator.  Technically this is
achieved by starting with a
trial value $\chi_{trial}$ and iteratively
solving Eq.~\ref{eq:linear} for
$\chi(t)$. On the order of 100 MD steps are
taken holding the pseudofermion
source fix, so that the operator A(t) changes
smoothly as a function of the MD
time t, as new values of U are generated.
These iterations, usually done by
the conjugate gradient (CG) method, are the
most computational expensive part
of Hybrid Monte Carlo algorithms.

This raises an obvious question. As we move in MD time,
why are we not
able to ``learn'' from the recent past enough
about the space of likely
solutions to vastly improve our iterative scheme?

One should be able to give a very good estimate,
$\chi_{trial}$ before
starting the conjugate gradient routine.
A crucial ingredient in this
approach, is the fact that detail balance
is in principle preserved
independent of the starting trial
configuration so long as one converges
accurately to the solution.
Therefore, we propose to estimate carefully the
starting configuration\cite{blk}.
To accomplish this some information on the
configurations in the previous MD
steps have to be stored. Although these
algorithms will use more memory,
memory is often not a severe constraint in
modern super computer simulations.

\section{ANALYTIC EXTRAPOLATIONS}

To motivate our extrapolation methods,
consider the function $\chi(t)$ as an
analytic function of t.  For simplicity
of notation suppose we want the value
at t = 0, given past values at $t_1, t_2, t_3, \cdots$.
In practice this is
usually a regular series of values $t_i = - i \; \epsilon$
with a integrations
step of size $dt = \epsilon$.
Then a trial value for the the solution,
$\chi(0)$ of the new Dirac matrix $A(0)$,
might be considered as a linear
superposition of old solutions,
\begin{equation}
       \chi_{trial} = c_1 \; \chi(t_1) + c_2 \; \chi(t_2)
        + \cdots + c_N \; \chi(t_N).
\end{equation}
If $N \epsilon$ is sufficiently small,
we may Taylor expand each term around t
= 0 and determine the coefficients by
canceling all terms for $\epsilon^k$ to
$O(\epsilon^N)$.
\begin{equation}
         \sum_{i=1}^{N} (t_i)^{n-1} \; c_i = \delta_{1,n}
\label{equ:taylor}
\end{equation}
As we will demonstrate this procedure is equivalent to the familiar N-1 order
polynomial fit to N points.

\subsection{Polynomial Extrapolation}

Good result can be obtained even with a si
mple polynomial fit of degree $N-1$.
To estimate the configuration it is necessary
to store in memory the previous
$(N+1)$ configurations.
However the polynomial extrapolation does not require
significant computational effort,
it is just a local sum on each lattice point
with fixed coefficient,
that is less than a single CG step.  The
$\chi_{trial}$ is expressed as a polynomial,
$\chi_{trial}~=~y_1 + t \; y_2 +\cdots + \;t^{N-1}\; y_N$,
whose coefficient satisfy the constraint,
\begin{equation}
      \sum_{n=1}^{N} (t_i)^{n-1} \; y_n = \chi(t_i) \; .
\label{equ:poly}
\end{equation}
One can easily prove that Eq.~\ref{equ:taylor} and
Eq.~\ref{equ:poly} define
identical extrapolation
($ y_1 = \sum_i c_i \chi(t_i) =\chi_{trial}$.  For
equally spaced time steps $t_i = - i \; \epsilon$,
the coefficients given by
\begin{equation}
       c_i~=~ (-1)^{i-1}~ {N! \over i! (N-i)!} \; .
\end{equation}
Table~\ref{tab:tavola1} shows the results
of simulations for polynomial
extrapolation. The number of CG steps needed
to reach convergence is plotted
in function of the degree of extrapolation $N$,
and the MD time step
$\epsilon$. The case $N=1$ corresponds
to starting with the old solution.

\begin{table*}[hbt]
\setlength{\tabcolsep}{1.2pc}
\newlength{\digitwidth} \settowidth{\digitwidth}{\rm 0}
\catcode`?=\active \def?{\kern\digitwidth}
\caption{CG steps using polynomial extrapolation}
\label{tab:tavola1}
\begin{tabular*}{\textwidth}{llllllll}
\hline
N=1 (previous data)
       & 0.98 & 0.93 & 0.93 & 0.88 & 0.85 & 0.77 & 0.63 \\
N=2 ($1^{st}$ order extrap.)
       & 0.92 & 1.07 & 0.86 & 0.74 & 0.72 & 0.62 & 0.27 \\
N=3 ($2^{nd}$ order extrap.)\
       & 0.91 & 0.82 & 0.74 & 0.61 & 0.59 & 0.47 & 0.21 \\
N=4 ($3^{th}$ order extrap.)
       & 0.96 & 0.77 & 0.56 & 0.44 & 0.41 & 0.31 & 0.21 \\
N=5 ($4^{th}$ order extrap.)
       & 0.86 & 0.77 & 0.54 & 0.45 & 0.42 & 0.35 & 0.26 \\
N=6 ($5^{th}$ order extrap.)
       & 0.70 & 0.48 & 0.39 & 0.38 & 0.40 & 0.40 & 0.41 \\
N=7 ($6^{th}$ order extrap.)
       & 0.70 & 0.59 & 0.40 & 0.42 & 0.42 & 0.39 & 0.46 \\
\hline
$\delta t=10^{-3}*$
       & 15   & 10   & 9    & 8    & 7    & 5    & 2    \\
\hline
\multicolumn{8}{@{}p{160mm}}{ Number of CG steeps
to reach the solution (residue $<~10^{-12}$),
normalized to 1 for $\chi=0$.
Full QCD configurations on $16^4$ lattice with
Wilson fermions $k=0.157$ and $\beta=5.6$.
Average on 30 configurations,
statistical errors are of the order of 6\%.  }
\end{tabular*}
\end{table*}

\begin{table*}[hbt]
\setlength{\tabcolsep}{1.2pc}
\catcode`?=\active \def?{\kern\digitwidth}
\caption{CG steps using minimum residual extrapolation}
\label{tab:tavola2}
\begin{tabular*}{\textwidth}{llllllll}
\hline
N=1 \ \ \ \ \ \ \ \ \ \ \ \ \ \ \ \ \ \ \ \ \ \ \ \ \
        & 0.99 & 0.93 & 0.92 & 0.87 & 0.84 & 0.81 & 0.64 \\
N=2     & 0.87 & 0.72 & 0.71 & 0.69 & 0.63 & 0.49 & 0.25 \\
N=3     & 0.85 & 0.64 & 0.57 & 0.52 & 0.49 & 0.32 & 0.06 \\
N=4     & 0.72 & 0.49 & 0.41 & 0.34 & 0.28 & 0.16 & 0.10 \\
N=5     & 0.72 & 0.37 & 0.31 & 0.23 & 0.19 & 0.12 & 0.07 \\
N=6     & 0.51 & 0.28 & 0.26 & 0.21 & 0.16 & 0.13 & 0.06 \\
N=7     & 0.49 & 0.25 & 0.24 & 0.21 & 0.19 & 0.12 & 0.06 \\
N=8     & 0.40 & 0.26 & 0.23 & 0.18 & 0.15 & 0.09 & 0.06 \\
N=9     & 0.38 & 0.23 & 0.19 & 0.16 & 0.13 & 0.09 & 0.06 \\
N=10    & 0.39 & 0.22 & 0.19 & 0.14 & 0.12 & 0.10 & 0.04 \\
N=11    & 0.38 & 0.18 & 0.17 & 0.14 & 0.11 & 0.08 & 0.04 \\
\hline
$\delta t=10^{-3}*$  & 15 & 10 & 9 & 8 & 7 & 5 & 2 \\
\hline
\multicolumn{8}{@{}p{160mm}}{ Number of CG steeps
to reach the solution (residue $<~10^{-12}$),
normalized to 1 for $\chi=0$.
Same configurations of Table 1.
Statistical errors are of the order of 4\%.}
\end{tabular*}
\end{table*}

\subsection{ Minimum Residual Extrapolation}

An alternate, perhaps more appealing, approach
is to consider the
past history of solutions,
$\chi(t_i)$ as defining an important
linear subspace for seeking an optimal trial solution.
Since the Conjugate
Gradient method is in fact just a minimal
residual technique confined
to the Krylov subspace spanned by
vectors $A^{j-1} \chi_{trial}$, why not
start by looking at a ``smarter''
subspace based on past success for
nearby times?

In this spirit, we suggest minimizing
the norm of the residual,
\begin{equation}
         r^\dagger r   =  \chi^\dagger M^\dagger M \chi -
        \varphi^\dagger \chi -  \chi^\dagger \varphi
        + b^\dagger b \; ,
\end{equation}
in the subspace spanned by $\chi_i \equiv \chi(t_i)$,
where $r = b - M \chi$ and $b \equiv M^\dagger \varphi$.
The minimization
condition reduces to
\begin{equation}
  \sum_{j=1}^{N}~{ \chi_i }^\dagger M^\dagger M \chi_j \; c_j =
   { \chi_i }^{\dagger} \varphi  \; .
\end{equation}
The only problem is that this system can be
ill conditioned because the
past solutions $ \chi_i $ differ from each
other by order $ \epsilon $.
Nevertheless if we only want to get
the minimum of $r^\dagger r$
in {\it span}($\chi_i$) using a
Gram Schmitt orthonormalization, we
can solve the system avoiding the singularities.

This method requires $(N^2+5N)/2$ dots product
and $(N)$ $M\chi$ matrix-vector
applications, and the storage of
$(N)$ past configuration.  It is
interesting to note that this
method gives coefficients that for the first
few order reproduce very close to the polynomial
extrapolation.
Table 2 show the number of CG steps using this method.

\section{CONCLUSIONS}

To compare efficiencies, the CG
iterations should be divided by $\epsilon$, so we
compare the total number of CG
steps needed to evolve the system for a fixed
distance in configuration space.
Note that if $\epsilon$ is too large, the
overall performance is good, but
the acceptance will drop drastically.  If
$\epsilon$ is too small, the extrapolation
is excellent, but the system will
evolve too slowly in phase space.
It is not trivial that we find a window
in $\epsilon$ where both the acceptance
is good and the extrapolation works
well.  Moreover this window has parameters
close to those used in actual
simulations.

Our present approach is clearly not the
only one worth considering. In fact we
have emphasized the analytic properties
of $\chi(t)$ because it suggests ways
to understand and further improve our approach.
For example the failure at
fixed $\epsilon$ to improve the polynomial
extrapolation by increasing
indefinitely the number of terms is probably
a signal of nearby singularities
in t. Our success so far is probably due
to low eigenvalues of the Dirac
operator changing more slowly with time.

Many other ways of using the past to
avoid needless repetition can be
imagined\cite{sant}. For example the
CG routine itself generates vectors, that
may be useful than solutions in the more distant past.
A Karman filter can be
used top introduce exponential decreasing
information from the past without
increasing the storage requirement.
The subspace of past vectors might be
used not just as a way to arrive at an initial guess,
but also as a way to
precondition the iterative process itself.
For now, we are pleased for now
that even a few vectors from the past
can be combined with simple
extrapolation ideas to give a
very useful acceleration method.

\end{document}